\documentclass[fleqn,12pt,twoside]{article}
\usepackage{espcrc1,epsfig}

\usepackage{graphicx}
\usepackage[figuresright]{rotating}

\newcommand{\AmS}{{\protect\the\textfont2
  A\kern-.1667em\lower.5ex\hbox{M}\kern-.125emS}}
\newcommand{\nc}{\newcommand}

\nc{\2}{\frac{1}{2}}
\nc{\half}{{\textstyle\2}}
\nc{\3}{\frac{1}{3}}
\nc{\4}{\frac{1}{4}}

\nc\be{\begin{equation}}
\nc\ee{\end{equation}}
\nc{\beq}[1]{\begin{equation}\label{#1}}
\nc{\eeq}{\end{equation}}
\nc{\bea}[1]{\begin{eqnarray}\label{#1}}
\nc{\eea}{\end{eqnarray}}

\nc{\bce}{\begin{center}}
\nc{\ece}{\end{center}}
\nc{\bit}{\begin{itemize}}
\nc{\eit}{\end{itemize}}
\nc{\bmp}{\begin{minipage}}
\nc{\emp}{\end{minipage}}
\nc{\bb}{\bm{b}}
\nc{\bq}{\bm{q}}
\nc{\bK}{\bm{K}}
\nc{\br}{\bm{r}}
\nc{\bs}{\bm{s}}

\nc{\Eq}{{\,=\,}}
\nc{\Kt}{K_\perp}
\nc{\pt}{p_\perp}
\nc{\mt}{m_\perp}
\nc{\pL}{p_{\rm L}}
\nc{\ET}{E_{\rm T}}
\nc{\Nch}{N_{\rm ch}}
\nc{\Nc}{N_{\rm coll}}
\nc{\Np}{N_{\rm part}}
\nc{\Atanh}{{\rm Atanh}}
\nc{\Asinh}{{\rm Asinh}}
\nc{\Acosh}{{\rm Acosh}}
\nc{\scm}{\sqrt{s_{\rm NN}}}

\nc{\la}{\langle}
\nc{\lla}{\left \langle}
\nc{\ra}{\rangle}
\nc{\rra}{\right \rangle}

\title{Emission angle dependent HBT at RHIC and beyond}

\author{Peter F. Kolb\address{Department of Physics and Astronomy,
                              SUNY at Stony Brook, NY 11794-3800, USA}
        and
        Ulrich Heinz\address{Physics Department, 
                        The Ohio State University, Columbus, OH 43210, USA}}
\begin{document}

\maketitle

\begin{abstract}
We study the geometrical features of non-central heavy ion collisions
throughout their dynamical evolution from equilibration 
to thermal freeze-out within a hydrodynamic picture. 
We discuss resulting observables, in particular 
the emission angle dependence of 
the HBT radii and the relation of these oscillations 
to the geometry at the final stage.
\end{abstract}


\section{Introduction}

The systematic investigation of observables generated by the
broken azimuthal symmetry of non-central nuclear collisions 
\cite{v2experiments}
has lead to major insight in the dynamics of the expansion.
In particular the large signal of 
anisotropic particle flow \cite{O92}
requires a surprisingly efficient microscopic rescattering 
mechanisms \cite{microv2} and even agrees with the limit 
of infinitely short rescattering lengths, namely
hydrodynamic calculations \cite{Kolbv2,Teaneyv2,Hirano},
thus providing access to the nuclear equation of state
under extreme conditions.
Furthermore, the large signal is critically dependent on
rapid thermalization and early pressure in the system,
at timescales smaller than 1 fm/$c$  \cite{KSH00}.
Experimentally and theoretically even more challenging is the
extraction and interpretation of the azimuthal dependence of 
coordinate space observables such as two-particle 
correlations \cite{HBTexp,W97,HK02HBT,HHLW02}.
Its understanding however offers an additional 
handle to study the space-time geometry of the source,
helps to understand and clarify persisting problems \cite{HK02}, 
and ultimately is one more piece in our effort to understand 
the {\em complete} dynamics of heavy ion collision.


\section{Fireball initialization and evolution}

To elucidate the connection of these observables and the
geometry that they reflect, we study two scenarios in the following. 
In the first one (referred to as `RHIC1') 
we choose the initial and freeze-out conditions
such that the particle yield as well as the 
pion and antiproton spectral slopes agree well with experimental data from  
{\em central} Au+Au collisions at 130 $A$GeV 
(for details, see \cite{Kolbv2}).
The maximum temperature in the center of the collision region
is found to be $T_0=340$~MeV  at an assumed equilibration time of 
$\tau_0=0.6$~fm/$c$.
The spectral slopes require a decoupling temperature of  $T_{\rm dec}=130$~MeV,
where fluid elements are assumed to liberate their particle content boosted by 
the local flow velocity.
{\em Geometry}, without involving further parameters,
then determines the initialization of {\em non-central} collisions.
Experimental results on elliptic flow confirmed the large 
values expected from hydrodynamic predictions \cite{KSH00}
and justified this idealized approach as long as the
produced fireball is sufficiently large to allow for 
thermalization and `macroscopic behavior', a requirement which
expectedly breaks down above a certain impact parameter.
For Au+Au collisions at $\sqrt{s_{\rm NN}}=130$ GeV 
the data indicate that this limit is reached for $b \sim 7-8$ fm 
\cite{Kolbv2}. 

With the second scenario we want to stress the geometrical features
of the source and the imprints on experimental observables. To enhance
the relevant signals, we employ more extreme conditions of $T_0=2$~GeV 
at $\tau_0=0.1$~fm/$c$ and $T_{\rm dec}=100$~MeV. As the system stays above
the QCD transition temperature for most of the time of the evolution we 
employ an ideal gas equation of state for simplicity
(for reasons that will become clear later we label this system `IPES').
%
%
At the moment it is not clear whether 
these conditions are accessible at future colliders, but the calculation 
allows us to  trace back the origin of qualitative changes 
in the observables which might already occur at LHC energies.
In the following we study the evolution of both systems for an 
impact parameter of 7 fm on the basis of an ideal hydrodynamic evolution with
longitudinal boost-invariance \cite{KSH00}. 

\begin{figure}[tbp]
\begin{minipage}[t]{80mm}
\epsfig{file=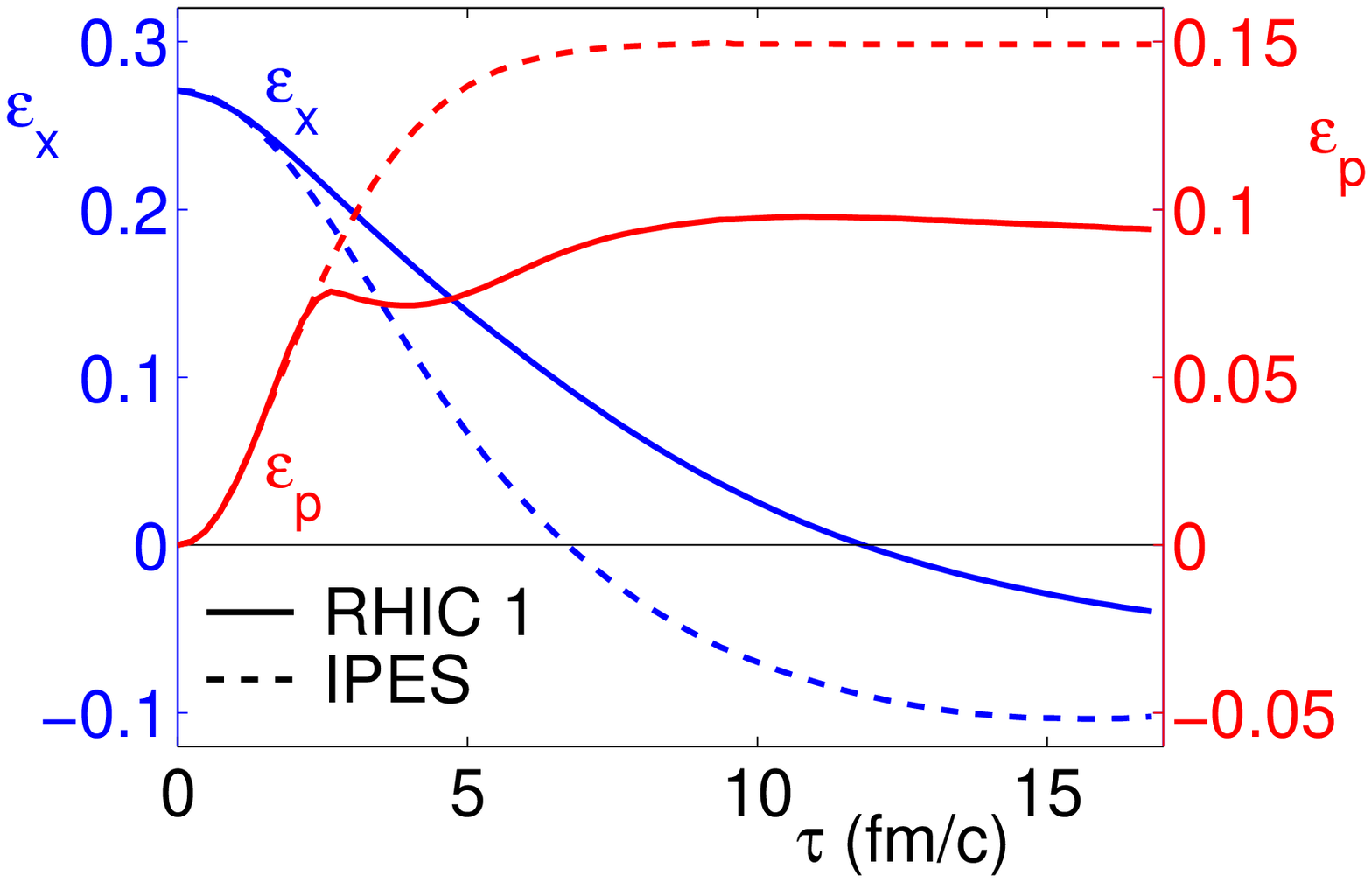, width=8.2cm}
\\[-1.5cm]
\caption{Time evolution of the spatial eccentricity $\epsilon_x$ 
         and momentum anisotropy $\epsilon_p$ for RHIC1 (solid)
         and IPES (dashed).}
\label{fig:epsevo}
\end{minipage}
\hspace{.7cm}
\begin{minipage}[t]{72mm}
\hspace*{2mm}
\epsfig{file=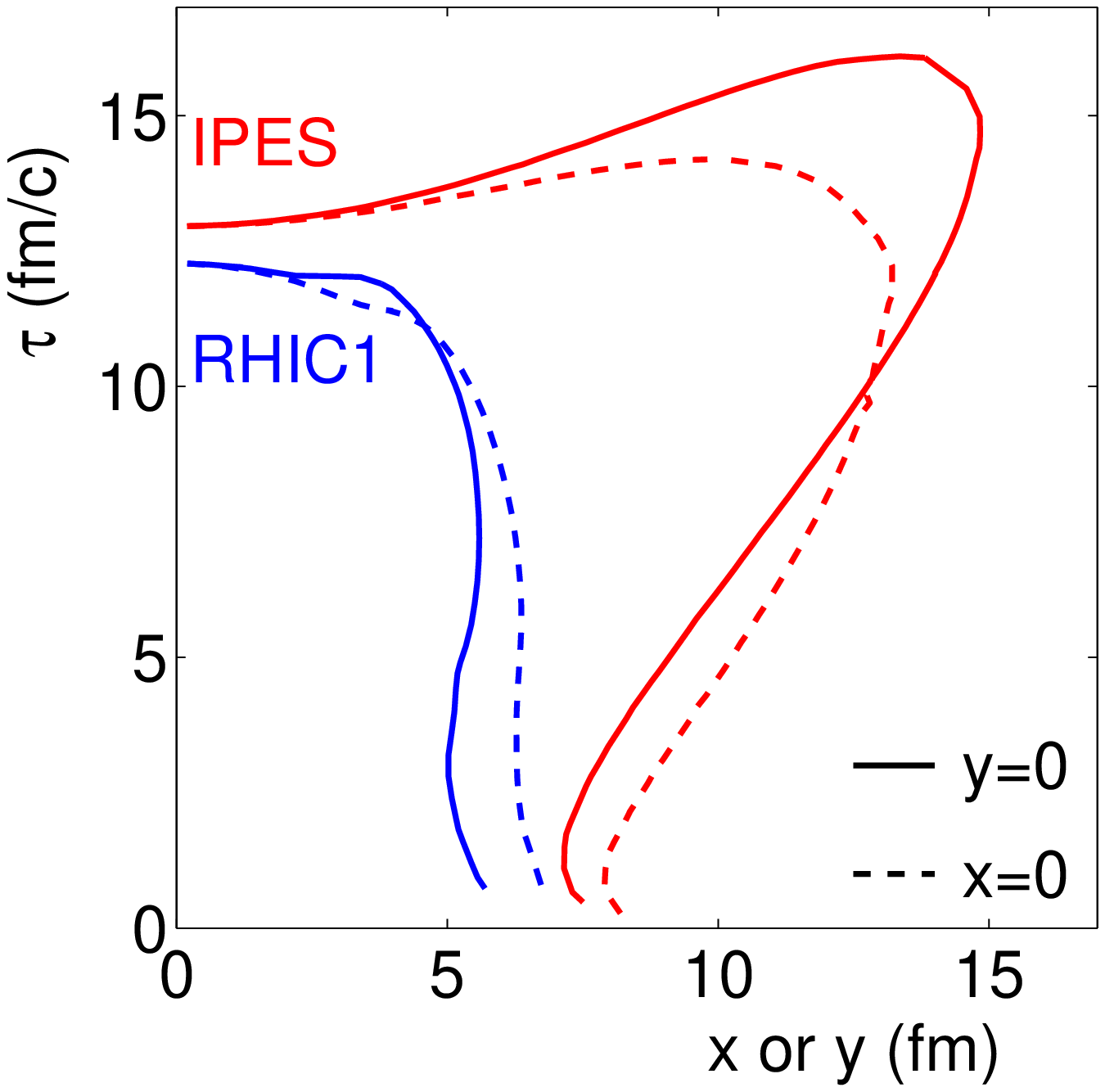, width=5.5cm,clip=}
\\[-1.5cm]
\caption{Freeze-out hypersurface of the two systems. 
         Shown are cuts at $y=0$ (solid) and $x=0$ (dashed).}
\label{fig:FOHS}
\end{minipage}
\\[-5mm]
\end{figure}

Upon impact the nuclei produce secondaries and large amounts of entropy 
in their geometrical overlap region. This is generally modeled by 
contributions from wounded nucleons and/or binary collisions to the
initial energy- or entropy density fields in the overlap region \cite{Kolbv2}. 
We characterize the spatial eccentricity of the fields by averaging over 
the transverse plane (at $z=0$) with the energy density as weight. 
Defining $\epsilon_x=\langle y^2-x^2 \rangle / \langle y^2 + x^2 \rangle$,
we can follow its evolution through time (see Fig. \ref{fig:epsevo}).
If the impact parameter points in $x$-direction, $\epsilon_x$ is positive
during the initial stages when the overlap-ellipsoid is elongated out of 
the reaction plane. 
As time evolves,
the stronger pressure gradients in the short direction force a stronger
expansion in plane than out of plane, thereby
reducing the initial anisotropy and eventually leading to an 
overshoot such that the source appears in plane elongated at 
the very end of its existence ($\epsilon_x < 0$).
This is in particular true for the `IPES' case, which exhibits much stronger 
pressure gradients and therefore stronger transverse expansion
(`IPES' stands for 'In Plane Elongated Source').
The generation of the accompanying anisotropic flow is 
characterized through the momentum anisotropy 
$\epsilon_p= \langle T^{xx}-T^{yy} \rangle / \langle T^{xx}+T^{yy} \rangle$
where the $T^{ii}$ are the diagonal elements of the 
energy momentum tensor of the fluid. Ultimately, this momentum anisotropy
is reflected as elliptic flow. Fig. \ref{fig:epsevo} shows clearly how the
geometrical eccentricity causes a rapid buildup of momentum anisotropy 
(the early saturation of $\epsilon_p$ at about 3 fm/$c$ in the RHIC1 case
is caused by the mixed phase of the employed equation of state \cite{KSH00}).

As particle emission occurs throughout the evolution from 
the surface of the fireball, it is not clear what average 
geometry will be imprinted on the particle distributions. 
Particles with high transverse momenta preferentially emerge 
from regions with the largest collective transverse velocities, 
that is from the freeze-out hypersurface at its largest radial 
extension, 
whereas the low $\pt$ particles are emitted when the interior
of the fireball freezes out in a timelike manner
(see the freeze-out hypersurfaces in Fig. \ref{fig:FOHS}). 
Therefore particles  of different $\pt$
(or particle pairs with different mean transverse momentum $K_\perp$
 as they are studied in HBT)
probe the system at different times and different regions
of the fireball.

\section{Azimuthally sensitive HBT-analysis}

From the hydrodynamically determined freeze-out hypersurface $\Sigma$
we evaluate the density of emitted pions of momentum $K$ through
$S(x,K) \propto
         \int_\Sigma K \cdot d^3 \sigma (x')   \,
                     f_{\rm BE}(E(x'),T_{\rm dec})\times$
                     $\delta^4(x-x')$
with $E(x')=K\cdot u(x')$ where $u$ is the local flow velocity.
With this source function $S(x,K)$ as a weight
we can calculate the average emission points and variances and
collect them in the spatial correlation tensor
$S_{\mu \nu} = \langle x_\mu x_\nu \rangle - \langle x_\mu \rangle
                                              \langle x_\nu \rangle$.
Rotating this correlation tensor into the direction of the particle
pair momentum (angle $\Phi$), one obtains the `width' and `depth'
of the emission region with respect to the direction of observation and can make the 
connection  \cite{W97} to the experimentally deduced HBT radii \cite{HBTexp}.

The `sideward' radius receives contributions only from the 
$(x,y)$-components of the correlation tensor and thus can be related
directly to the geometry of the fireball. More specifically it reflects 
the `width' of the emission region of particles of similar momentum
orthogonal to their average momentum (the direction of observation).
For large transverse momenta, these `homogeneity regions' are narrow 
slivers close to the rim of the freeze-out hypersurface whose shape is
quite different from that of the total source \cite{HK02HBT}. 
The `out' and `out-side' terms mix geometrical and temporal emission 
information, with the temporal part gaining weight with increasing 
transverse momentum. In the calculations we can investigate both parts 
separately, but experimentally they appear always in combination.

\begin{figure}[tbp]
\begin{minipage}[t]{75mm}
\epsfig{file=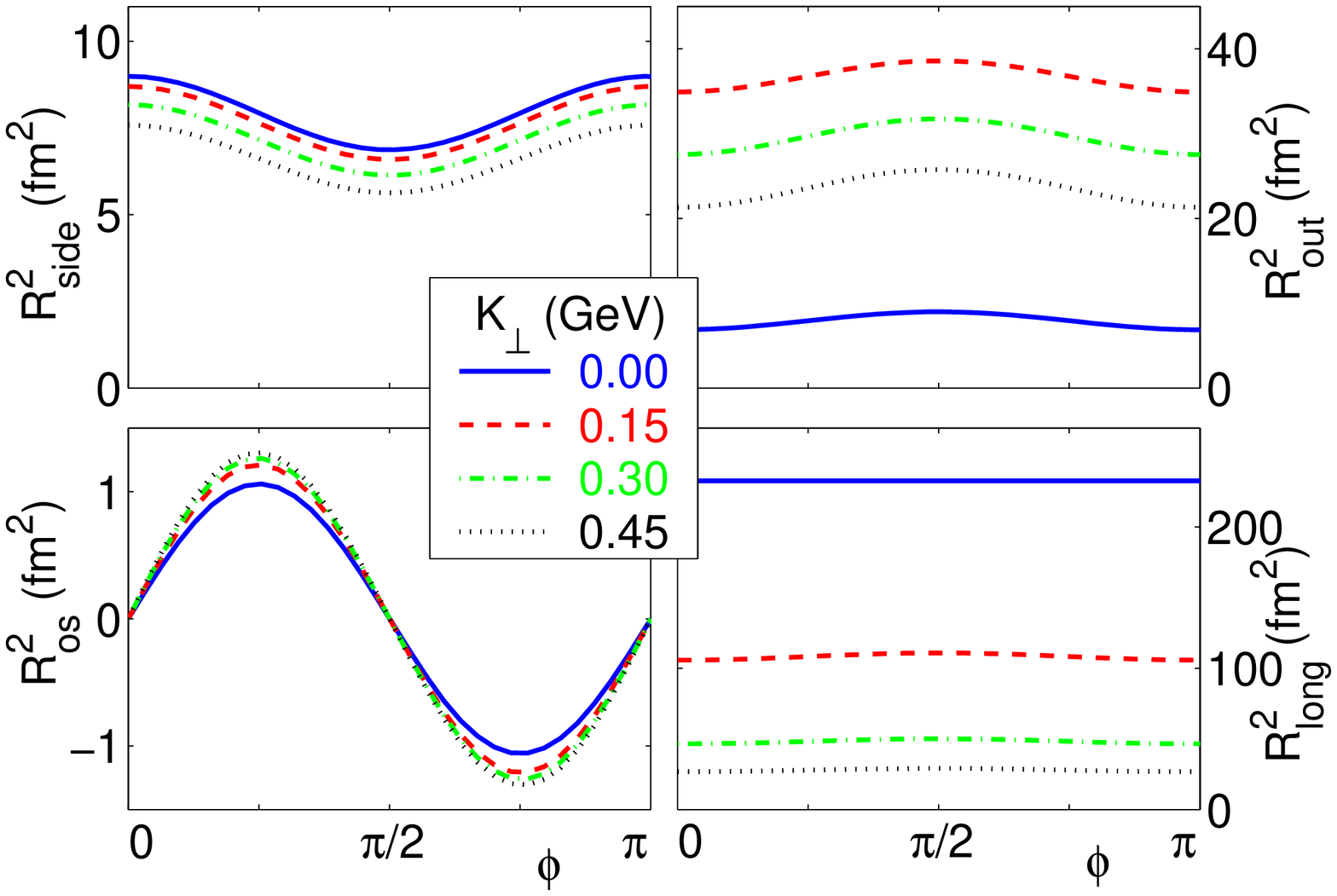, width=7.5cm}
\\[-15mm]
\caption{Oscillations of the sideward, outward, out-side and
longitudinal radius for the RHIC1 system.}
\label{fig:RHICHBT}
\end{minipage}
\hspace{\fill}
\begin{minipage}[t]{75mm}
\epsfig{file=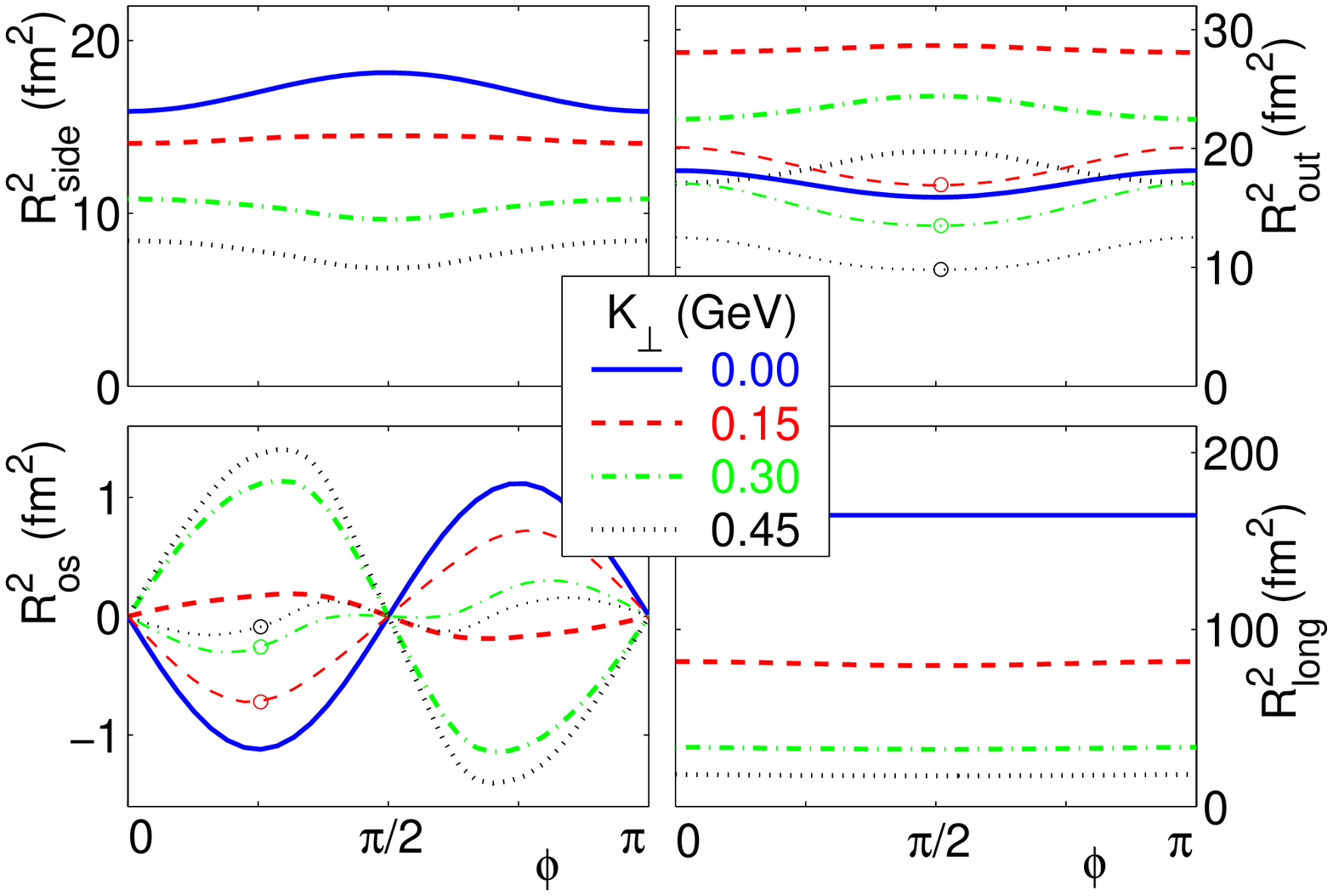, width=7.5cm}
\\[-15mm]
\caption{As Fig. \ref{fig:RHICHBT} for the IPES system.
         The purely geometrical contributions are marked by circles.
}
\label{fig:IPESHBT}
\end{minipage}
\\[-5mm]
\end{figure}

Fig. \ref{fig:RHICHBT} displays the oscillations of the squared radii for 
the RHIC1 calculations. Each radius shows its own oscillation pattern 
which remains qualitatively unchanged for all transverse momenta. The 
sideward oscillation reflects the {\em out-of-plane} extended source: 
observing such a source from the $x$-direction it appears to be larger 
sideways than when observed from the $y$-direction. This remains true
even at large $K_\perp$ where the emission is concentrated along a thin
sliver near the rim of the fireball \cite{HK02HBT}. The outward 
oscillations confirm this behavior from an 'orthogonal' viewpoint. 
The sign of the out-side oscillation can be traced back to a 'tilt' 
of the emission region relative to the emission direction \cite{HK02HBT} 
which is generated by the out-of-plane deformation of the source.
Even though hydrodynamic calculations fail in the quantitative description 
of the freeze-out geometry \cite{HK02}, the overall source shape appears to 
be in agreement with preliminary data \cite{HBTexp}.

Moving to the IPES calculation a drastic change occurs 
(Fig.~\ref{fig:IPESHBT}): The oscillations flip sign at $K_\perp 
 \sim 0.15$~GeV, showing the same oscillation pattern as at RHIC1 for 
larger $K_\perp$ but opposite oscillations below. The geometric 
contributions to the out and out-side radius (marked by circles) always 
oscillate opposite to the RHIC1 case, reflecting the in-plane elongation 
of this source. At larger $K_\perp$ this naively expected behavior
is masked by growing temporal contributions, leading eventually
to the same qualitative oscillation pattern as seen for RHIC1. Because 
the homogeneity regions do not trace out the interior of the source
but concentrate near the rim of the system, the oscillations of the
sideward radius also change sign at larger $K_\perp$. The true geometry 
of the source is therefore only reflected in the correlations of the 
particles with the smallest transverse momenta. 

{\bf Acknowledgements:} 
This work was supported in part by the U.S. Department of Energy under 
Contracts No. DE-FG02-88ER40388 and  DE-FG02-01ER41190.
PFK is supported by a Feodor Lynen fellowship 
of the Alexander von Humboldt Foundation.

\end{document}